# Microscopic quantum ideal triaxial rotor model and related self-consistent cranking model: slow-wobbling rotation in $^{20}_{10}Ne$

P. Gulshani

NUTECH Services, 3313 Fenwick Crescent, Mississauga, Ontario, Canada L5L 5N1
Tel. #: 647-975-8233; matlap@bell.net

A microscopic quantum ideal rotor-model intrinsic Hamiltonian for triaxial rotation is derived from the nuclear Schrodinger equation by applying a rotation operator to a deformed nuclear ground state. This Hamiltonian is obtained only when a rigid-flow prescription is used for the three rotation angles in the rotation operator. Using Hartree-Fock variational and second quantization methods, the rotor Hamiltonian is transformed into that of a self-consistent triaxial cranking model (MSCRM-3) with a microscopically and self-consistently defined angular-velocity vector, plus residual terms associated with the square of the angular momentum operator and with a two-body interaction. The approximations and assumptions underlying the conventional cranking model are revealed. For a self-consistent deformed harmonic oscillator potential, the MSCRM-3 Schrodinger equation is transformed into that of a uniaxial cranking model plus local potential-energy cross terms using a rotation of the co-ordinate system. It is shown that uniform rotation, where the angular-momentum vector is aligned with the angular-velocity vector, is not generally possible. However, for a slow-wobbling rotation, an approximate uniform rotation becomes possible. In this limiting wobbly motion, the potential-energy cross terms are negligibly small, and the uni-axial cranking-model equation is solved analytically using a generalization of the isotropic-velocity-distribution condition of Bohr-Mottelson and Ripka-Blaizot-Kassis. The ground-state rotational-band excitation energy and quadrupole moment for the slow-wobbling rotation are calculated and compared with the measured data in $^{20}_{10}Ne$. The results provide a physical explanation for the hitherto mysterious decrease in the excitation-energy level spacing with increasing angular momentum, and for the sharp drop in the quadrupole moment near the band termination in $^{20}_{10}Ne$. The impact of the residual of the square of the angular momentum and a separable quadrupole-quadrupole two-body interaction is studied in the Tamm-Dancoff approximation using the eigenstates of the self-consistent cranking model as the particle-hole basis states.

## 1. Introduction

In reference [1], the phenomenological semi-classical self-consistent conventional cranking model for a rotation about a single axis (CRM1) was derived from first principles. The derivation restored the time-invariance in CRM1 and clarified the assumptions and approximations that underlie CRM1, and identified corrections to the model. In particular, it was shown that CRM1 Hamiltonian is purely intrinsic (i.e., independent of the rotation angle and angular momentum) only if the rotation angle is determined by a rigid-flow velocity field.

CRM1 has successfully explained and predicted many of the observed rotational features in deformed nuclei that can be modeled by a rotation about a single principal axis of the deformed nuclear shape. Other nuclear rotational features (such as precession, wobbling, etc.), which involve rotation about an arbitrary axis, have been explained [2-18] using the following simple triaxial generalization (CRM3) of CRM1 Hamiltonian:

$$\hat{H}_{cr} \cdot |\Phi_{cr}\rangle \equiv \left(\hat{H}_o - \omega_1^{cr} \cdot \hat{J}_1 - \omega_2^{cr} \cdot \hat{J}_2 - \omega_3^{cr} \cdot \hat{J}_3\right) \cdot |\Phi_{cr}\rangle = E_{cr} |\Phi_{cr}\rangle \tag{1}$$

where the nuclear Hamiltonian $\hat{H}_o$ is:

$$\hat{H}_o = \sum_{n=1}^{A} \frac{\hat{p}_n^2}{2M} + \frac{1}{2} \sum_{n,m}^{A} \hat{V}\left(|\vec{r}_n - \vec{r}_m|\right) \tag{2}$$

$\hat{V}$ (spin, isospin, and exchange dependence of $\hat{V}$ is left out for now) is a rotationally invariant two-body interaction, $A$ is the mass number, and $\vec{\omega}^{cr}$, with constant Cartesian components $\omega_k^{cr}$ ($k = 1, 2, 3$), is the rotation or cranking angular-velocity vector $\vec{\omega}^{cr}$.
However, like CRM1, CRM3 is phenomenological and semi-classical in nature. In particular, the nucleus is externally, rather than self-) driven in the model (i.e., $\vec{\omega}^{cr}$ is an arbitrary constant vector) and $\hat{H}_{cr}$ is not rotationally and time-reversal invariant. It is desirable to derive CRM3 from first principles and reveal the underlying approximations and assumptions and any corrections to it, as we have done for CRM1 in [1]. There have been such studies [19-48] using angular momentum projection, generator co-ordinate, density matrix, etc. methods, and using various approximations and assumptions such as large deformations, expansion in powers of the angular momentum operator, and using lowest-order density matrices.

In this article, we generalize the uniaxial-rotation model in [1] to a microscopic quantum ideal triaxial rotor model for rotation about all the three axes of the deformed nuclear shape, which is then reduced to a self-consistent triaxial cranking model plus residual microscopic quantum corrections.

In Section 2 of this article, we derive from first principles an ideal triaxial rotor model (that is distinct from that of Bohr's rotational model) by applying a rotation operator to a deformed nuclear ground-state (or excited rotational-band band-head) wavefunction and thereby obtain the corresponding rotor nuclear Hamiltonian. Each of the three rotation angles in the rotation operator is chosen to be defined by a rigid-flow component of the nucleon velocity field. This choice eliminates terms in the transformed nuclear Hamiltonian that would otherwise couple the angular momentum to other operators, giving the rotor Hamiltonian a purely (ideally) intrinsic or non-rotational character and containing only the square of the angular momentum. For the center-of-mass motion, the method predicts the correct mass.

In Section 3, we use Hartree-Fock (HF) mean-field variational approximation to reduce the rotationally and time-reversal invariant ideal triaxial rotor model of Section 2 to a self-consistent triaxial cranking model plus the remaining part (i.e., the HF direct and exchange parts of the one-body and two-body parts) of the square of the angular-momentum operator and two-body interaction, which are then treated as a residual interactions. Approximations and assumptions underlying the conventional cranking model are identified.

In Section 4, the HF self-consistent triaxial cranking model Schrodinger equation for a self-consistent deformed harmonic oscillator potential is transformed to that of a uniaxial cranking model equation plus local potential cross terms using a co-ordinate system rotation. In the limit of a slow wobbly rotation about a single axis, the angular velocity about this axis becomes much larger than the transverse angular velocities and the effect of the local potential cross terms becomes negligibly small. Then, the uniaxial cranking-model equation is solved analytically using a generalization of the isotropic-velocity-distribution condition of Bohr-Mottelson [49] and Ripka-Blaizot-Kassis [50], and the ground-state rotational-band excitation energy and quadrupole moment are determined.

In Section 5, we present the predictions of the self-consistent cranking model for slow wobbling rotation and its comparison with the measured data for $^{20}_{10}Ne$.

In Section 6, we solve the microscopic ideal rotor-model Schrodinger equation, including the residual angular-momentum operator and a residual schematic two-body interaction, in the Tamm-Dancoff approximation using the cranked HF states derived in Section 4 as particle-hole basis states.

In Section 7, the microscopic ideal rotor model of Section 6 is used to predict the ground-state rotational-band excitation energy and quadrupole moment in $^{20}_{10}Ne$ and the results are compared with the corresponding measured data.

Section 8 concludes the article.

## 2. Derivation of microscopic quantum ideal rotor model for 3-D rotation

To derive the microscopic quantum ideal rotor model (which is distinct from Bohr's rotational model) for a triaxial rotation, we consider a deformed nuclear ground state described by a wavefunction $|\Phi_{gs}\rangle$ obtained by some method such as HF. We assume that $|\Phi_{gs}\rangle$ is an approximate ground state of the nucleus and hence approximately satisfies the nuclear Schrodinger equation for a rotationally-invariant Hamiltonian $\hat{H}_o$:

$$\hat{H}_o |\Phi_{gs}\rangle = E_{gs} |\Phi_{gs}\rangle \qquad (3)$$

where the nuclear Hamiltonian $\hat{H}_o$ is defined in Eq. (2). Next, we rotate $|\Phi_{gs}\rangle$ (i.e., the deformed nucleon density distribution) through an angel $\theta$ about the axis of the rotation-angle vector $\vec{\theta} = \theta \cdot \vec{e}$, where $\vec{e}$ is a unit vector along $\vec{\theta}$, with Cartesian components $\theta_k$ ($k = 1, 2, 3$) along space-fixed frame axes, to obtain a rotated state $|\Phi(\theta_k)\rangle$ as follows[1]:

$$|\Phi(\theta_k)\rangle = e^{i\vec{\theta}\cdot\hat{\vec{J}}/\hbar} \cdot |\Phi_{gs}\rangle \equiv e^{Z} \cdot |\Phi_{gs}\rangle \tag{4}$$

The rotationally-invariant rotation operator in Eq. (4) is similar to that in [51] and transforms the Hamiltonian symmetrically along the three axes, unlike the rotation vector in terms of the Euler angles used in the angular-momentum projection and other studies [8,23,24,52,53,54]. The rotational invariance of the rotation operator in Eq. (4) simplifies the transformation of the Hamiltonian and eliminates the need to specify the action of the angular momentum on the angles in this transformation (in contrast to the situation in the uniaxial-rotation case in [1]).

Substituting Eq. (4) for $|\Phi_{gs}\rangle$ into Eq. (3), we obtain the transformed Schrodinger equation:

$$\hat{H}|\Phi(\theta_k)\rangle = E_{gs}|\Phi(\theta_k)\rangle \tag{5}$$

where the transformed nuclear Hamiltonian $\hat{H}$ is given by:

$$\hat{H} \equiv e^{Z} \cdot \hat{H}_o \cdot e^{-Z} = \frac{1}{2M}\sum_{n=1}^{A} e^{Z} \cdot \hat{p}_n^2 \cdot e^{-Z} + \frac{1}{2}\sum_{n,m}^{A} \hat{V}(|\vec{r}_n - \vec{r}_m|) \tag{6}$$

and where we have used the rotational invariance of $\hat{V}$. Since $\vec{\theta}$ in Eq. (4) defines the orientation of a deformed nuclear state and this orientation changes from a state to state according to Eq. (5), we conclude that $\vec{\theta}$ is a dynamical variable and is therefore considered to be a function of at least the nucleon co-ordinates. Therefore, in Eq. (6), $\vec{\theta}$ and $\hat{p}_n^2$ do not commute. Eq. (5) shows that each of the rotated wavefunction $|\Phi(\theta_k)\rangle$ in Eq. (4) satisfies Eq. (5) with the same ground-state energy $E_{gs}$. In other words, the Hamiltonian $\hat{H}$ and $|\Phi(\theta_k)\rangle$ for all orientations $\theta_k$ describe degenerate (or collapsed) rotational states with the same energy $E_{gs}$. Hence, we may conclude that $\hat{H}$ has an intrinsic character (which is examined further below) and $|\Phi(\theta_k)\rangle$ is a superposition of angular momentum eigenstates. (For the case of the center-of-mass motion, Eq. (6) becomes: $\hat{H} = H_o - \frac{1}{2MA}P^2$, where $\vec{P}$ is the center-of-mass momentum, and hence the correct mass $MA$ is predicted.) The transformed Hamiltonian $\hat{H}$ in Eq. (6) is observed to be rotationally invariant since:

$$[\vec{J}, \hat{H}] = [\vec{J}, e^{Z}\hat{H}e^{-Z}] = e^{Z}[\vec{J}, \hat{H}_o]e^{-Z} = 0 \tag{7}$$

where we have used the rotational invariance of $Z$. The invariance in Eq. (7) is one of the major advantages of the rotation operator $e^{Z}$ over other forms of the rotation operator such as the successive Euler rotation operators used in the literature.

---

[1] We use the space-fixed-frame axis components of the rotation vector $\vec{\theta}$ because it simplifies the analysis and provides a transformed $\hat{H}_o$ in terms of operators along the space-fixed frame axes, whereas the usual Euler rotation operator $e^{i\alpha\cdot\hat{J}_3/\hbar} \cdot e^{i\beta\cdot\hat{J}_2/\hbar} \cdot e^{i\gamma\cdot\hat{J}_3/\hbar}$ in terms of the Euler angles along body-fixed frame axes and the corresponding operator $e^{i\theta_1\cdot\hat{J}_1/\hbar} \cdot e^{i\theta_2\cdot\hat{J}_2/\hbar} \cdot e^{i\theta_3\cdot\hat{J}_3/\hbar}$ obtained from the Euler rotation operator have complicated transformation properties.

We evaluate $e^{Z} \cdot \hat{p}_n^2 \cdot e^{-Z}$ in Eq. (6) using the following expansion in powers of Z:

$$e^{Z} \cdot \hat{p}_n^2 \cdot e^{-Z} = \hat{p}_n^2 + \left[Z, \hat{p}_n^2\right] + \frac{1}{2!}\left[Z,\left[Z, \hat{p}_n^2\right]\right] + \ldots \ldots \quad (8)$$

Using the rotational invariance of Z and $\hat{p}_n^2$ in Eq. (7), we can easily evaluate the terms in Eq. (8) and show that the terms in higher powers of Z vanish. Substituting these results into Eq. (6), we obtain:

$$\hat{H} = \hat{H}_o + \frac{i\hbar}{2M} \cdot \sum_{n=1}^{A}\sum_{k=1}^{3} \nabla_n^2 \theta_k \cdot \hat{J}_k - \frac{1}{M} \cdot \sum_{n=1}^{A}\sum_{k=1}^{3} \vec{\nabla}_n \theta_k \cdot \vec{\hat{p}}_n \cdot \hat{J}_k$$
$$+ \frac{1}{2M} \cdot \sum_{n=1}^{A}\sum_{k=1}^{3} \vec{\nabla}_n \theta_k \cdot \vec{\nabla}_n \theta_l \cdot \hat{J}_k \cdot \hat{J}_l \quad (9)$$

To ensure zero coupling between $\vec{J}$ and other (such as shear) operators in the third term on the right-hand-side of Eq. (9) coming from the $\vec{\hat{p}}_n$ term, we use, as in [1], the following rigid-flow prescription for the rotation angles $\vec{\theta}$:

$$\frac{\partial \theta_1}{\partial x_{nj}} = -\sum_{k=2}^{3} {}_1\chi_{jk} \, x_{nk}, \qquad {}_1\chi_{jk} = -{}_1\chi_{kj} = 0 \text{ for } j,k \neq 2,3 \quad (10)$$

$$\frac{\partial \theta_2}{\partial x_{nj}} = -\sum_{k=1,\neq 2}^{3} {}_2\chi_{jk} \, x_{nk}, \qquad {}_2\chi_{jk} = -{}_2\chi_{kj} = 0 \text{ for } j,k \neq 1,3 \quad (11)$$

$$\frac{\partial \theta_3}{\partial x_{nj}} = -\sum_{k=1}^{2} {}_3\chi_{jk} \, x_{nk}, \qquad {}_3\chi_{jk} = -{}_3\chi_{kj} = 0 \text{ for } j,k \neq 1,2 \quad (12)$$

where ${}_k\chi$ is a real $3 \times 3$ anti-symmetric matrix. Each of the selections in Eqs. (10), (11), and (12) picks the collective rigid-flow component of the velocity field of each nucleon, rendering the third term on the right-hand-side of Eq. (9) quadratic in $\vec{J}$. The non-zero elements of ${}_k\chi$ are determined by choosing $\theta_k$ and $\hat{J}_l$ to be canonically conjugate, i.e.,

$$\left[\theta_k, \hat{J}_l\right] = i\hbar \delta_{kl} \quad (13)$$

Substituting Eqs. (10), (11), and (12) into Eq. (13), we obtain:

$${}_1\chi_{kl} = \frac{1}{\hat{\mathscr{I}}_1^+}, \qquad {}_2\chi_{kl} = -\frac{1}{\hat{\mathscr{I}}_2^+}, \qquad {}_3\chi_{kl} = \frac{1}{\hat{\mathscr{I}}_3^+} \quad (14)$$

$$\hat{\mathscr{I}}_1^+ \equiv \sum_{n=1}^{A}\left(y_n^2 + z_n^2\right), \quad \hat{\mathscr{I}}_2^+ \equiv \sum_{n=1}^{A}\left(x_n^2 + z_n^2\right), \quad \hat{\mathscr{I}}_3^+ \equiv \sum_{n=1}^{A}\left(x_n^2 + y_n^2\right) \quad (15)$$

where $M\hat{\mathscr{I}}_k^+$ is the $k^{th}$ principal-axis component of the rigid-flow moment of inertia tensor[2]. We observe that, for the choice in Eqs. (10), (11), (12), (14), and (15), we obtain the result:

---

[2] Note that the rigid-flow prescription for $\theta_k$ in Eqs. (10), (11), (12), and (14) is a collective analogue of the Birbrair's single-particle $\theta_n$ [55]: $\vec{\nabla}_n \theta_n = \vec{e}_x \times \vec{r}_n / (y_n^2 + z_n^2)$, where $\vec{e}_x$ is a unit vector along the x axis. $\theta_n$ has continuous second-order mixed derivatives (i.e., $\vec{\nabla}_n \times \vec{\nabla}_n \theta_n = 0$) in any spatial region that excludes the x axis along which $\theta_n$ is singular. Whereas $\theta_k$ has discontinuous second-order mixed derivatives (i.e., $\vec{\nabla}_n \times \vec{\nabla}_n \theta_k \neq 0$). The difference in the discontinuity between $\vec{\nabla}_n \theta_k$ and $\vec{\nabla}_n \theta_n$ arises because of the many-body nature of $\left(\hat{\mathscr{I}}_k^+\right)^{-1}$ in

$$\left[\theta_k, \hat{J}_l\right] = i\hbar \frac{Q_{kl}}{\hat{\mathcal{S}}_k^+} \approx 0 \ (k \neq l) \tag{16}$$

since $Q_{lk}$ for $k \neq l$ is very small compared to $\hat{\mathcal{S}}_k^+$, where $Q_{lk} \equiv \sum_{n=1}^{A} x_{nl} \cdot x_{nk}$ is the $kl^{th}$ component of the quadrupole moment tensor. (However, there is no need to know the action of $\hat{J}_l$ on $\theta_k$ for $k \neq l$ in the derivations of the equations in this article.) Substituting Eqs. (10), (11), (12), (14), and (15) into Eq. (9), we obtain:

$$\hat{H} \equiv \hat{H}_o - \sum_{k=1}^{3} \frac{1}{2M\hat{\mathcal{S}}_k^+} \cdot \hat{J}_k^2 - \sum_{k \neq l=1}^{3} \frac{Q_{lk}}{2M\hat{\mathcal{S}}_l^+ \hat{\mathcal{S}}_k^+} \cdot \hat{J}_l \cdot \hat{J}_k \tag{17}$$

We observe that the second term on the right-hand side of Eq. (17) resembles the rotational kinetic energy of a rotating triaxial classical rigid body [56] with the difference that all the operators are space-fixed-frame (and not body-fixed) axis components. The third term in Eq. (17) is negligibly small because $M^2 \hat{\mathcal{S}}_l^+ \hat{\mathcal{S}}_k^+$ is very large and the off-diagonal elements $MQ_{lk}$ of the quadrupole moment tensor are small, and hence the third term can be and is neglected.

Inserting Eq. (17) into Eq. (5) and neglecting the last term in Eq. (17), we obtain the following Schrodinger equation for rotation of the microscopic quantum ideal triaxial rotor:

$$\hat{H}\left|\Phi(\vec{\theta})\right\rangle \equiv \left[\hat{H}_o - \sum_{k=1}^{3} \frac{\hat{J}_k^2}{2M\hat{\mathcal{S}}_k^+}\right]\left|\Phi(\vec{\theta})\right\rangle = E_{Jgs}\left|\Phi(\vec{\theta})\right\rangle \tag{18}$$

where now we have added the subscript $J$ to $E_{gs}$ to indicate its implicit $J$ dependence.

We may consider the time-reversal and rotationally invariant (within the approximation used in Eq. (17)) Hamiltonian $\hat{H}$ in Eq. (18) to be an intrinsic Hamiltonian determining a degenerate deformed rotated (or rotating) intrinsic state $\left|\Phi(\vec{\theta})\right\rangle = e^Z \left|\Phi_{gs}\right\rangle$ with the orientation $\vec{\theta}$ for a system with the kinematic moments of inertia $M\hat{\mathcal{S}}_k^+$. Indeed, we observe in Eq. (18) that the rotational kinetic energy is subtracted from the nuclear Hamiltonian $\hat{H}_o$, leaving us with the intrinsic Hamiltonian $\hat{H}$[3]. In fact, one can easily show that[4]:

---

Eq. (14). However, the discontinuity in $\vec{\nabla}_n \theta_k$ is small because the expectation of $\hat{\mathcal{S}}_k^+$ is a large number. On the other hand, for a rigid-body motion $\vec{\nabla}_n \theta_{rig} = \omega_{rig} \vec{e}_x \times \vec{r}_n$, with a constant angular velocity $\omega_{rig}$, $\vec{\nabla}_n \theta_{rig}$ is inherently discontinuous. We note that second-order mixed derivatives of $\theta_k$ do not appear anywhere in the derivation of the equations in this article.

[3] In many studies, the operator $\sum_{k=1}^{3} \hat{J}_k^2 / \mathcal{S}_k$ with some inertia parameters $\mathcal{S}_k$ has been used to remove spurious rotational-excitation energy. The above results then show that for this removal to be exactly valid, $\mathcal{S}_k$ must be replaced by $M\hat{\mathcal{S}}_k^+$. A Hamiltonian of the form of $\hat{H}$ in Eq. (18) with an arbitrary inertia parameters in place of $M\hat{\mathcal{S}}_k^+$ has been used in many studies of nuclear collective rotation [36-45,57-61].

[4] The conditions in Eqs. (7) and (19) are similar to those for a Goldstone boson or phonon arising in an RPA mode with zero excitation energy and zero restoring force, and identified with a rotational motion [39,41-45].

$$\left[\theta_k, \hat{H}\right] = 0 \tag{19}$$

The result in Eq. (19) proves that $\hat{H}$ in Eq. (18) is a purely or ideally intrinsic Hamiltonian. It follows that the wavepacket $|\Phi(\theta)\rangle$ in Eqs. (4), (5), and (18) is not an eigenstate of the angular momentum, but rather is a superposition of such states. As in the conventional cranking model in [1], we require $|\Phi(\theta)\rangle$ to describe a state with a given average angular momentum satisfying the condition:

$$\hbar^2 J(J+1) = \langle \Phi | \hat{J}^2 | \Phi \rangle = \sum_{k=1}^{3} \langle \Phi | \hat{J}_k^2 | \Phi \rangle \tag{20}$$

Eq. (18) can be expressed as follows:

$$\hat{H}_o |\Phi(\vec{\theta})\rangle = \left[\hat{H} + \sum_{k=1}^{3} \frac{\hat{J}_k^2}{2M \hat{\mathcal{S}}_k^+}\right] |\Phi(\vec{\theta})\rangle \equiv \overline{E}_J |\Phi(\vec{\theta})\rangle \tag{21}$$

Eq. (21) resembles that in the Bohr's rotational model [22,24,29,32,52,57-63] but differs inherently from it in the following ways: $|\Phi\rangle$ is not a product of an angular momentum eigenstate and an intrinsic state but rather is a superposition of angular momentum eigenstates (i.e., it is not a product of rotation and intrinsic wavefunctions as in the Bohr's rotational model), and $M\hat{\mathcal{S}}_k^+$ are kinematic rather than dynamic moments of inertia even though $\hat{H}$ an intrinsic Hamiltonian as in the Bohr's model. Therefore, Eq. (21) and Bohr's corresponding equation are not expected to predict similar results. In fact, from an analysis of angular momentum projection of deformed HF states, Bouten-Caurier [63] have concluded that the rotational motion predicted by the Bohr's rotational model is not applicable to the rotational motion in light nuclei, except at very large deformation, whereas the conventional cranking model has been largely successful in predicting rotational spectra in the light nuclei. The above results and the success of the cranking model in predicting nuclear rotational properties imply that a clean and complete separation of rotation and intrinsic motions in nuclei may not be possible (except at very large deformation). This conclusion seems to be also supported by the analyses in [33,34,35] where it is found that such a separation is not possible except under an extreme condition.

It may be of interest to note that Eq. (21) may be related to the procedure used by Skyrme-Levinson-Kelson, et. al., [57-60] in computing an approximate average value of the dynamic moment of inertia.

## 3. Deriving HF cranking model plus residual operators from Eq. (18)

To reduce Eq. (18) to that of a self-consistent version of the conventional cranking model in Eq. (1) plus any remaining correction terms, we first make the many-body operator $\left(\hat{\mathcal{S}}_k^+\right)^{-1}$ in Eq. (18) manageable by replacing $\hat{\mathcal{S}}_k^+$ in Eq. (18) by its expectation:

$$\mathcal{S}_k^o \equiv \langle \Phi | \hat{\mathcal{S}}_k^+ | \Phi \rangle \tag{22}$$

Eq. (18) then becomes:

$$\hat{H}\left|\Phi\left(\vec{\theta}\right)\right\rangle \equiv \left(\hat{H}_o - \sum_{k=1}^{3}\frac{\hat{J}_k^2}{2M\mathcal{S}_k^o}\right)\left|\Phi\left(\vec{\theta}\right)\right\rangle = E_{Jgs}\left|\Phi\left(\vec{\theta}\right)\right\rangle \tag{23}$$

We now apply to the expectation of the Hamiltonian $\hat{H}$ in Eq. (23) HF variational and second quantization methods as described in [1], where we include in the HF variation only the direct HF term $\left\langle \hat{J}_k \right\rangle^2$ of $\left\langle \hat{J}_k^2 \right\rangle$, and include the remaining parts of $\left\langle \hat{J}_k^2 \right\rangle$ in the residual $\left(\hat{J}_k^2\right)_{res}$ of $\hat{J}_k^2$ (refer to Eq. (29) below). The HF variation then yields the following single-particle quantum self-consistent cranking-model equation[5]:

$$\hat{h}_{crHF}\left|\varphi_n\right\rangle \equiv \left(\hat{h}_{oHF} - \vec{\Omega}\cdot\vec{\hat{j}}\right)\left|\varphi_n\right\rangle = \varepsilon_n\left|\varphi_n\right\rangle \tag{24}$$

where $\hat{h}_{oHF}$ is the HF mean-field part of $\hat{H}_o$:

$$\hat{h}_{oHF} \equiv \frac{1}{2M}\hat{p}^2 + v(\vec{x}) \tag{25}$$

$v$ is the direct HF mean-field single-particle local potential part of $\hat{V}$ in Eq. (2), $\vec{\Omega}\cdot\vec{\hat{j}}$ is the single-particle direct HF mean-field part of $\sum_{k=1}^{3}\frac{\hat{J}_k^2}{2M\mathcal{S}_k^o}$, and where the three Cartesian components $\Omega_k$ of the rotation angular-velocity vector $\vec{\Omega}$ are defined by:

$$\Omega_k = \left\langle \Phi_{crHF}\left|\hat{J}_k\right|\Phi_{crHF}\right\rangle / M\mathcal{S}_k^o \tag{26}$$

where $\mathcal{S}_k^o$ is given in Eq. (22), and the $A$-particle cranked HF ground state $\left|\Phi_{crHF}\right\rangle$ is defined by:

$$\left|\Phi_{crHF}\right\rangle \equiv \hat{P}\prod_{n=1}^{A}\left|\varphi_n\right\rangle \tag{27}$$

and $\hat{P}$ is the anti-symmetrization operator. The angular momentum constraint on the wavefunction in Eq. (20) then becomes:

$$\hbar^2 J(J+1) = \left\langle \Phi_{crHF}\left|\hat{J}^2\right|\Phi_{crHF}\right\rangle = \sum_{k=1}^{3}\left\langle \Phi_{crHF}\left|\hat{J}_k^2\right|\Phi_{crHF}\right\rangle = \sum_{k=1}^{3}\left\langle \Phi_{crHF}\left|\hat{J}_k\right|\Phi_{crHF}\right\rangle^2 \tag{28}$$

where the third equality in Eq. (28) is valid because $\left|\Phi_{crHF}\right\rangle$ is a HF state. Eqs. (26) and (28) determine the $J$-dependence of $\vec{\Omega}$. The rigid-flow condition in Eq. (26) ensures that the Hamiltonian $\hat{H}$ in Eq. (23) is purely an intrinsic quantity as defined in Eq. (19). This condition is in addition to that required by the HF mean-field approximation (such as minimization of the mean-field energy subject to a constant volume when an approximate potential is used as a substitute for the actual HF mean-field potential).

---

[5] In this variation, the contribution from the change in $\mathcal{S}_k^o$ is neglected because it is small compared to that from the expectation of $\hat{J}_k^2$ since $\mathcal{S}_k^o$ is large and varies little with $J$ or $\vec{\Omega}$. For these reasons, we have replaced $\hat{\mathcal{S}}_k^+$ in Eq. (21) by $\mathcal{S}_k^o = \left\langle \Phi\left|\hat{\mathcal{S}}_k^{+o}\right|\Phi\right\rangle$ to obtain Eq. (23).

The microscopic quantum self-consistent cranking model for triaxial rotation Eqs. (24) and (26) become identical to the conventional cranking model Eq. (1) when the kinematic moment of inertia $M\mathfrak{S}_k^o$, or equivalently the rotation angular velocity $\vec{\Omega}$, in Eqs. (24) and (26) is replaced by an arbitrary inertia parameter. But then the Hamiltonian $\hat{H}$ in Eq. (23) would no longer be a purely intrinsic quantity. Therefore, the conventional cranking-model Hamiltonian in Eq. (1) is not a purely intrinsic Hamiltonian because it does not use the rigid-flow condition in Eq. (26). The conventional and microscopic self-consistent cranking model ignore the HF exchange term of the one-body part and other residual parts of the square of the angular momentum operator.

To regain the full Hamiltonian $\hat{H}$ in Eq. (23), we must add to the HF mean-field in Eq. (24), the remaining or residual parts of the nuclear interaction $\hat{V}$ in $\hat{H}_o$ (refer Eq. (2)) and $\hat{J}_k^2$ of Eq. (23). The residual part of $\hat{J}_k^2$ is given in the second-quantized representation by:

$$\left(\hat{J}_i^2\right)_{res} \equiv \hat{J}_i^2 - 2\langle \Phi_{crHF} | \hat{J}_i | \Phi_{crHF} \rangle \cdot \hat{J}_i = \sum_{\nu,\nu'=1}^{\infty}\left[\left(\hat{j}_i^2\right)_{\nu\nu'} - 2\sum_{k=1}^{A}\left(\hat{j}_i\right)_{\nu k}\cdot\left(\hat{j}_i\right)_{k\nu'}\right]\cdot a_\nu^\dagger a_{\nu'}$$
$$+ \sum_{l,k=1}^{A}\left(\hat{j}_i\right)_{lk}\cdot\left(\hat{j}_i\right)_{kl} - \langle \hat{J}_i\rangle^2 + \sum_{\mu,\mu',\nu,\nu'=1}^{\infty}\left(\hat{j}_i\right)_{\mu\mu'}\cdot\left(\hat{j}_i\right)_{\nu\nu'}:a_\mu^\dagger a_\nu^\dagger a_{\nu'} a_{\mu'}: \quad (29)$$

where $\hat{j}_i$ is the $i^{th}$ component ($i=1,2,3$) of the single-particle angular momentum operator, the subscripts $k$ and $l$ and $\mu$ and $\nu$ indicate matrix elements between the (HF) self consistent cranked orbitals in Eq. (24), angled brackets indicate expectation over cranked HF nuclear ground state, and colons indicate normal ordering with respect to the cranked HF ground state $|\Phi_{crHF}\rangle$. The first two terms in Eq. (29) are the direct and exchange HF parts of the one-body part of $\hat{J}_k^2$. The next term in Eq. (29) is the HF expectation of one-body part of $\hat{J}_k^2$. The last term in Eq. (29) is the residual of the two-body part of $\hat{J}_k^2$.

We now add to the HF mean-field Schrodinger Eq. (24) the residual part $\hat{V}_{res}$ of the two-body interaction $\hat{V}$ in $\hat{H}_o$ in Eq. (2), and the residual part of the rotational kinetic energy in Eq. (23). Eq. (24) then becomes:

$$\hat{H}|\Phi\rangle = \left[\hat{H}_{crHF} + \hat{V}_{res} - \sum_{k=1}^{3}\frac{\left(\hat{J}_k^2\right)_{res}}{2M\mathfrak{S}_k^o}\right]|\Phi\rangle = E_{Jgs}|\Phi\rangle \quad (30)$$

where

$$\hat{H}_{crHF}|\Phi_{crHF}\rangle \equiv \sum_{n=1}^{A}\left[\hat{h}_{oHF}(n) - \vec{\Omega}\cdot\hat{\vec{j}}(n)\right]|\Phi_{crHF}\rangle = \left[\hat{H}_{oHF} - \vec{\Omega}\cdot\hat{\vec{J}}\right]|\Phi_{crHF}\rangle = E_{JgscrHF}|\Phi_{crHF}\rangle \quad (31)$$

$$E_{JgscrHF} \equiv \sum_{n=1}^{A}\varepsilon_n \quad (32)$$

where $\varepsilon_n$ and the $A$-particle HF ground state $|\Phi_{crHF}\rangle$ are given in Eqs. (24) and (27) respectively, and $\langle \hat{J}_k^2 \rangle_{res}$ is given in Eq. (29). For $\hat{V}_{res}$ we choose the separable effective quadrupole-quadrupole residual interaction (which is often used in nuclear structure calculations):

$$\hat{V}_{res} = -\frac{\chi}{2} Q^\dagger \cdot Q, \qquad \hat{Q}_\mu \equiv \sum_{n=1}^A r_n^2 \hat{q}_n^\mu \equiv \sum_{n=1}^A r_n^2 \cdot Y_{2\mu}(\theta_n, \phi_n), \qquad \mu = 0, \pm 1, \pm 2 \qquad (33)$$

where the parameter $\chi$ is the interaction strength. Eq. (30) is equivalent to the microscopic quantum ideal rotor Hamiltonian in Eq. (23) in the HF representation.

## 4. Solving self-consistent cranking-model Eqs. (24) and (26)

In this section, we solve the microscopic self-consistent cranking-model Schrodinger Eq. (24) (or Eq. (31)) with the rigid-flow kinematic moment of inertia constraint in Eq. (26) using a self-consistent deformed harmonic oscillator potential for $v$ in Eq. (25):

$$\hat{H}_{oHF} = \sum_{n=1}^A \hat{h}_{oHF} = \frac{1}{2M} \cdot \sum_{n,j=1}^{A,3} \hat{p}_{nj}^2 + \frac{M\omega_1^2}{2} \cdot \sum_n x_n^2 + \frac{M\omega_2^2}{2} \cdot \sum_n y_n^2 + \frac{M\omega_3^2}{2} \cdot \sum_n z_n^2, \qquad (34)$$

To make the oscillator potential in Eq. (34) resemble a HF mean-field potential (i.e., satisfy HF self-consistency and nuclear force short-range features), the ground-state energy $E_{JgscrHF}$ in Eq. (32) is minimized with respect to the oscillator frequencies $\omega_k$ subject to the constant nuclear potential volume condition[6]:

$$\omega_1 \cdot \omega_2 \cdot \omega_3 = \omega_o^3 \qquad (35)$$

where the isotropic harmonic-oscillator potential frequency $\omega_o$ is given by $\hbar\omega_o = P \cdot A^{-1/3}$, and $P = 41$ for heavy and 35.43 for light nuclei [22,24,49,50, 52,64,65]. This minimization yields a self-consistency between the shapes of nuclear equi-potential and equi-density surfaces [49,50,64-67].

The cranking model Eq. (24) (or (31)) for a deformed-harmonic-oscillator potential in Eq. (34) and for an arbitrary rotational angular velocity $\vec{\Omega}$ has been solved numerically [3,4,6,10,13]. We solve Eq. (24) by reducing it to the equation of the cranking model for a rotation about a single axis. In other words, we reduce Eq. (24) to a case of uniform rotation about a tilted-axis, suggested in [6,10]. We do so by transforming Eqs. (24) and (34) to a rotated co-ordinate system whose $x$-axis coincides with the rotation vector $\vec{\Omega}$:

$$\vec{\Omega} \equiv (\Omega_1, \Omega_2, \Omega_3) \equiv \Omega(\cos\theta \cdot \cos\phi, \cos\theta \cdot \sin\phi, \sin\theta) \qquad (36)$$

This transformation is achieved by rotating the reference frame $(x,y,z) \equiv (x_1, x_2, x_3)$ counter-clockwise through the angle $\phi$ about the $z$-axis, then clock-wise through the angle $\theta$ about the new $y$-axis, and finally through the angle $\gamma$ about the new $x$-axis to obtain the co-ordinate system $(x',y',z') \equiv (x'_1, x'_2, x'_3)$ given in terms of two $3 \times 3$ real orthogonal matrices $R$ and $\bar{R}$ defined as follows:

$$x_k = \sum_{l,j=1}^{3} R_{kl} \cdot \bar{R}_{lj} \, x'_j, \quad R \equiv \begin{pmatrix} \cos\phi \cdot \cos\theta & -\sin\phi & -\cos\phi \cdot \sin\theta \\ \sin\phi \cdot \cos\theta & \cos\phi & -\sin\phi \cdot \sin\theta \\ \sin\theta & 0 & \cos\theta \end{pmatrix}, \quad \bar{R} \equiv \begin{pmatrix} 1 & 0 & 0 \\ 0 & \cos\gamma & -\sin\gamma \\ 0 & \sin\gamma & \cos\gamma \end{pmatrix} \quad (37)$$

---

[6] The condition in Eq. (35) is a simple, reasonably accurate approximation to the usually used constant ellipsoid volume condition: $\langle x^2 \rangle \cdot \langle y^2 \rangle \cdot \langle z^2 \rangle = c_o$ where $\langle x_k^2 \rangle \equiv \langle \Phi_{crHF} | \sum_n x_{nk}^2 | \Phi_{crHF} \rangle$ ($k = 1,2,3$) and $c_o$ is a constant.

From Eqs. (36) and (37), one can readily express the elements of $R$ in terms of the dimensionless frequencies:

$$\tilde{\Omega}_k \equiv \frac{\Omega_k}{\Omega}, \qquad \tilde{\Omega}_1^2 + \tilde{\Omega}_2^2 + \tilde{\Omega}_3^2 = 1 \qquad (38)$$

Substituting the first of Eqs. (37) into the cranking model Hamiltonian $\hat{h}_{crHF}$ in Eq. (24), we obtain:

$$\hat{h}'_{crHF} |\varphi'_n\rangle \equiv \left[ \hat{h}'_{oHF} + M x' \cdot (\omega'^2_{12} y' + \omega'^2_{13} z') - \vec{\Omega} \cdot \hat{j}' \right] |\varphi'_n\rangle = \varepsilon_n |\varphi'_n\rangle \qquad (39)$$

where:

$$\hat{h}'_{oHF} \equiv \frac{1}{2M} \hat{p}'^2 + \frac{M \omega'^2_1}{2} x'^2 + \frac{M \omega'^2_2}{2} y'^2 + \frac{M \omega'^2_3}{2} z'^2 \qquad (40)$$

and where the matrix $\bar{R}$, i.e., the angle $\gamma$, is chosen such that the coefficient of the cross-term $y'z'$ vanishes. This condition yields:

$$\tan 2\gamma = \frac{2\omega'^2_{23}}{\omega'^2_2 - \omega'^2_3} \qquad (41)$$

In Eq. (39), $\hat{j}'$ is the component of $\vec{j}$ along the $x'$-axis, which coincides with $\vec{\Omega}$. The frequencies appearing in Eqs. (39) and (40) are defined as follows:

$$\omega'^2_1 \equiv \sum_{k=1}^{3} \tilde{\Omega}_k^2 \cdot \omega_k^2, \quad \omega'^2_2 \equiv \frac{\tilde{\Omega}_2^2 \cdot \omega_1^2 + \tilde{\Omega}_1^2 \cdot \omega_2^2}{\tilde{\Omega}_2^2 + \tilde{\Omega}_1^2},$$

$$\omega'^2_3 \equiv \frac{\tilde{\Omega}_1^2 \cdot \tilde{\Omega}_3^2}{\tilde{\Omega}_2^2 + \tilde{\Omega}_1^2} \omega_1^2 + \frac{\tilde{\Omega}_2^2 \cdot \tilde{\Omega}_3^2}{\tilde{\Omega}_2^2 + \tilde{\Omega}_1^2} \omega_2^2 + (\tilde{\Omega}_2^2 + \tilde{\Omega}_1^2) \omega_3^2 \qquad (42)$$

$$\omega'^2_{12} \equiv \frac{\tilde{\Omega}_1^2 \cdot \tilde{\Omega}_2^2}{\tilde{\Omega}_2^2 + \tilde{\Omega}_1^2}(\omega_1^2 - \omega_2^2), \quad \omega'^2_{13} \equiv -\frac{\tilde{\Omega}_3}{\tilde{\Omega}_2^2 + \tilde{\Omega}_1^2} \cdot (\tilde{\Omega}_1^2 \cdot \omega_1^2 + \tilde{\Omega}_2^2 \cdot \omega_2^2) + \tilde{\Omega}_3 \cdot \sqrt{\tilde{\Omega}_2^2 + \tilde{\Omega}_1^2} \cdot \omega_3^2,$$

$$\omega'^2_{23} \equiv \frac{\tilde{\Omega}_1 \cdot \tilde{\Omega}_2 \cdot \tilde{\Omega}_3}{\tilde{\Omega}_2^2 + \tilde{\Omega}_1^2} \cdot (\omega_1^2 - \omega_2^2) \qquad (43)$$

Eq. (39) resembles the cranking Hamiltonian for uniform rotation about a tilted axis in [6,10]. It also resembles that for a tilted-axis cranked harmonic oscillator in [4] but with a major difference: the frequencies $\omega'^2_{12}$, $\omega'^2_{13}$, and $\omega'^2_{23}$, associated with the off-diagonal elements of the rotated harmonic-oscillator-potential energy quadrupole tensor $q_{lk} \equiv x_l x_k$ for $l \neq k$ or cross terms in Eq. (39), are functions of the space-fixed frame frequencies $\omega_k^2$ and hence $\omega'^2_{12}$, $\omega'^2_{13}$, and $\omega'^2_{23}$ are not independent of each other and of $\omega_k^2$. On the other hand, in [4] the cross-term frequencies corresponding to $\omega_{12}'^2$, $\omega_{13}'^2$, and $\omega_{23}'^2$ are assumed to be independent parameters, and hence they are varied independently of $\omega_k^2$ to minimize the intrinsic energy. Thereby [4] concludes that the cross-term frequencies must vanish, which in turn requires a rotation about a principal axis (i.e., a uniform rotation). A related observation is that, in the microscopic cranking model Eq. (39), the rotation angular velocity $\vec{\Omega}$ is not an arbitrary parameter as in the conventional cranking model but rather it is determined by the rigid-flow constraint in Eq. (26) (which is needed to ensure that the Hamiltonian $\hat{H}$ in Eq. (23) is purely an intrinsic quantity as defined in Eq. (19)). Therefore, $\vec{\Omega}$ is not parallel to $\vec{J} \equiv \langle \Phi_{crHF} | \vec{\hat{J}} | \Phi_{crHF} \rangle$, and hence uniform rotation is not possible unless the three kinematic moments of inertia $\mathcal{S}_k^o$, defined in Eq. (22), are the same[7]. We show below that, ignoring the potential-energy cross-terms in Eq. (39), renders the three moments $\mathcal{S}_k^o$ the same. Ignoring the cross terms reduces the Hamiltonian in Eq. (39) to that for a uniaxial principal-axis rotation with the additional feature of a small-amplitude precession or wobbling of $\vec{J}$, as is shown below.

From Eqs. (39), (41), and (43), we observe that the potential-energy cross-term frequencies $\omega_{12}'^2$, $\omega_{13}'^2$, and $\omega_{23}'^2$ are at least an order of magnitude smaller than the diagonal-term frequencies $\omega_k'^2$ in Eqs. (40) and (42) since $\tilde{\Omega}_k^2 \leq 1$ as a consequence of the second of Eqs. (38), and hence $\tilde{\Omega}_1^2 \cdot \tilde{\Omega}_2^2 \ll 1$ particularly for a small-amplitude or slow precession or wobbling rotation where one, say $\tilde{\Omega}_1$, of the three components of the rotation vector $\vec{\Omega}$ is large (close to unity) and the other two components $\tilde{\Omega}_2$ and $\tilde{\Omega}_2$ are small (much less than unity) (this is discussed further below). Therefore, we ignore the cross terms $x'y'$ and $x'z'$ in Eq. (39), which then reduces to:

$$\hat{h}_{crHF}' | \varphi_n' \rangle \equiv \left[ \hat{h}_{oHF}' - \vec{\Omega} \cdot \hat{\vec{j}}' \right] | \varphi_n' \rangle = \varepsilon_n | \varphi_n' \rangle \tag{44}$$

Eq. (44) is simply the Schrodinger equation for a uniaxial cranking model, and its solution has been determined [1,50,64-70] using the canonical or unitary transformation:

$$| \varphi_n' \rangle = U | \varphi_{n,os}' \rangle, \qquad U^\dagger \hat{h}_{crHF}' U | \varphi_{n,os}' \rangle = \varepsilon_n | \varphi_{n,os}' \rangle \tag{45}$$

to eliminate the cross terms $y' p_{z'}$ and $z' p_{y'}$ in Eq. (44), and obtain the following three independent (uncoupled) oscillators with the wavefunction $| \varphi_{n,os}' \rangle = | \varphi_{n_1,os}' \rangle | \varphi_{n_2,os}' \rangle | \varphi_{n_3,os}' \rangle$, the Hamiltonian:

$$\overline{\hat{h}}_{crHF}' \equiv U^\dagger \hat{h}_{crHF}' U = \frac{1}{2M} \hat{p}'^2 + \frac{M \omega_1'^2}{2} x'^2 + \frac{M \alpha_2'^2}{2} y'^2 + \frac{M \alpha_3'^2}{2} z'^2 \tag{46}$$

and the $A$-particle energy eigenvalue:

---

[7] Note that Eq. (44) and the conclusion drawn from it are valid for any definition of the rotation angle $\vec{\theta}$ in Eq. (4) and the corresponding kinematic moment of inertia $\mathcal{S}_k^o$ and not just for rigid-flow constraint in Eqs. (10), (11), (12).

$$E_{JgscrHF} = \sum_{k=1}^{3}\sum_{n_k=0}^{n_{fk}} \varepsilon_{n_k} = \hbar\omega_1'\Sigma_1 + \hbar\alpha_2'\Sigma_2 + \hbar\alpha_3'\Sigma_3 \tag{47}$$

where the intrinsic system frequencies ($\alpha_2'$, $\alpha_3'$) and the total oscillator particle occupation numbers $\Sigma_k$ are:

$$\alpha_2'^2 \equiv \omega_+'^2 + \Omega^2 + \sqrt{\omega_-'^4 + 4\Omega^2\omega_+'^2}, \quad \alpha_3'^2 \equiv \omega_+'^2 + \Omega^2 - \sqrt{\omega_-'^4 + 4\Omega^2\omega_+'^2}, \quad \Sigma_k \equiv \sum_{n_k=0}^{n_{kf}}(n_k + 1/2) \tag{48}$$

$$\omega_+'^2 \equiv (\omega_2'^2 + \omega_3'^2)/2, \quad \omega_-'^2 \equiv (\omega_2'^2 - \omega_3'^2)/2, \quad \Sigma_k \equiv \sum_{n_k=0}^{n_{kf}}(n_k + 1/2) \tag{49}$$

and $\omega_1'$ is given in Eq. (42), and $n_{kf}$ is the number of oscillator quanta in the $k^{\text{th}}$ direction at the Fermi surface.

Using the solution in Eq. (45), we determine the expectations of the components of the angular momentum operator and rigid-flow moments of inertia Eqs. (15) and (22):

$$\langle\Phi_{crHF}|\hat{J}_k|\Phi_{crHF}\rangle = \langle\Phi_{crHF}'|\hat{J}_k'|\Phi_{crHF}'\rangle = \hbar\Omega_k\mathcal{I}_o = \hbar\Omega_k \cdot \left[\mathcal{I}_{rig} - \frac{4}{\alpha_2^2 - \alpha_3^2}(\alpha_3\Sigma_3 - \alpha_2\Sigma_2)\right] \tag{50}$$

$$\mathcal{I}_3^o \equiv \langle\Phi_{crHF}|\hat{\mathcal{I}}_3^+|\Phi_{crHF}\rangle = \langle\Phi_{crHF}'|\hat{\mathcal{I}}_3'^+|\Phi_{crHF}'\rangle$$
$$= \frac{\hbar}{M} \cdot \left[(1-\tilde{\Omega}_3^2)\frac{\Sigma_1}{\omega_1'} + \frac{1+\tilde{\Omega}_3^2}{2}\mathcal{I}_{rig} + \frac{\omega_-^2}{\alpha_2^2-\alpha_3^2}\left(\frac{\Sigma_2}{\alpha_2} - \frac{\Sigma_3}{\alpha_3}\right) \cdot \frac{1-\tilde{\Omega}_3^2}{\sqrt{1+\tan^2 2\gamma}}\right] \tag{51}$$

$$\mathcal{I}_2^o \equiv \langle\Phi_{crHF}|\hat{\mathcal{I}}_2^+|\Phi_{crHF}\rangle = \langle\Phi_{crHF}'|\hat{\mathcal{I}}_2'^+|\Phi_{crHF}'\rangle$$
$$= \frac{\hbar}{M} \cdot \left[(1-\tilde{\Omega}_2^2)\frac{\Sigma_1}{\omega_1'} + \frac{1+\tilde{\Omega}_2^2}{2}\mathcal{I}_{rig}\right.$$
$$\left.+ \frac{\omega_-^2}{\alpha_2^2-\alpha_3^2}\left(\frac{\Sigma_2}{\alpha_2} - \frac{\Sigma_3}{\alpha_3}\right) \cdot \frac{1}{\sqrt{1+\tan^2 2\gamma}}\left(\frac{1+\tilde{\Omega}_3^2}{1-\tilde{\Omega}_3^2}\cdot\tilde{\Omega}_2^2 - 1 + \frac{2\Omega_1\Omega_2\Omega_3}{1-\tilde{\Omega}_3^2}\tan 2\gamma\right)\right] \tag{52}$$

$$\mathcal{I}_1^o \equiv \langle\Phi_{crHF}|\hat{\mathcal{I}}_1^+|\Phi_{crHF}\rangle = \langle\Phi_{crHF}'|\hat{\mathcal{I}}_1'^+|\Phi_{crHF}'\rangle$$
$$= \frac{\hbar}{M} \cdot \left[(1-\tilde{\Omega}_1^2)\frac{\Sigma_1}{\omega_1'} + \frac{1+\tilde{\Omega}_1^2}{2}\mathcal{I}_{rig}\right.$$
$$\left.+ \frac{\omega_-^2}{\alpha_2^2-\alpha_3^2}\left(\frac{\Sigma_2}{\alpha_2} - \frac{\Sigma_3}{\alpha_3}\right) \cdot \frac{1}{\sqrt{1+\tan^2 2\gamma}}\left(\frac{1+\tilde{\Omega}_3^2}{1-\tilde{\Omega}_3^2}\cdot\tilde{\Omega}_1^2 - 1 - \frac{2\Omega_1\Omega_2\Omega_3}{1-\tilde{\Omega}_3^2}\tan 2\gamma\right)\right] \tag{53}$$

where the $A$-particle cranked HF ground state $|\Phi_{crHF}\rangle$ is:

$$|\Phi_{crHF}\rangle = |\Phi'_{crHF}\rangle \equiv \hat{P}\prod_{n=1}^{A}|\varphi'_n\rangle \tag{54}$$

$$\mathcal{I}_o \equiv -\left[\frac{\Sigma_2}{\alpha_2} + \frac{\Sigma_3}{\alpha_3} + \frac{4\omega_+^2}{\alpha_2^2 - \alpha_3^2}\left(\frac{\Sigma_2}{\alpha_2} - \frac{\Sigma_3}{\alpha_3}\right)\right], \quad \mathcal{I}_{rig} \equiv \frac{\Sigma_2}{\alpha_2} + \frac{\Sigma_3}{\alpha_3} + \frac{4\Omega^2}{\alpha_2^2 - \alpha_3^2}\left(\frac{\Sigma_2}{\alpha_2} - \frac{\Sigma_3}{\alpha_3}\right) \tag{55}$$

where $\mathcal{I}_{rig}$ is the rigid-flow moment of inertia for a rotation without angular-momentum precession about a principal axis at angular velocity $\Omega$. Eqs. (50), (51), (52), and (53) show that precession (i.e., non-zero values of $\Omega_k$ ($k = 1,2,3$)) reduces the kinematic moments of inertia $\mathcal{I}_k^o$ ($k = 1,2,3$) and $\mathcal{I}_o$ below the rigid-flow value $\mathcal{I}_{rig}$ (this is further discussed in Section 5). Eq. (50) can also be expressed in the vector form:

$$\vec{J} \equiv \langle\Phi_{crHF}|\vec{\hat{J}}|\Phi_{crHF}\rangle = \hbar\vec{\Omega}\mathcal{I}_o \tag{56}$$

where $\mathcal{I}_o$ is defined in (55). Eq. (56) shows that $\vec{J}$ is parallel to the rotation vector $\vec{\Omega}$ because the three kinematic moments of inertia $\bar{\mathcal{I}}_k$ defined generally by (compare the rigid-flow constraint in Eq. (26)):

$$\langle\Phi_{crHF}|\hat{J}_k|\Phi_{crHF}\rangle = \hbar\Omega_k\bar{\mathcal{I}}_k \tag{57}$$

are the same (i.e., $\bar{\mathcal{I}}_k = \mathcal{I}_o$ for ($k = 1,2,3$)) as a consequence of ignoring the potential-energy cross terms $x'y'$ and $x'z'$ in Eq. (39), as is shown in the footnote 8[8], where it is shown that, with the cross terms present, uniform rotation in general not possible.

Substituting Eqs. (50) to (53) into the rigid-flow constraint in Eq. (26) (which ensures that the Hamiltonian $\hat{H}$ is purely intrinsic, refer to Section 2 and Eqs. (10), (11), (12), and (15)), we obtain:

$$\hbar\Omega_k \cdot \left[\mathcal{I}_{rig} - \frac{4}{\alpha_2^2 - \alpha_3^2}(\alpha_2\Sigma_2 - \alpha_3\Sigma_3)\right] = M\Omega_k\mathcal{I}_k^o = M\Omega_k\mathcal{I}_o \tag{58}$$

Eq. (58) determines $\Omega_k$'s.

The rotation angular velocity $\Omega$ in Eq. (55) is determined from the angular momentum constraint in Eq. (28), which in view of the result in Eq. (56), becomes:

$$\hbar^2 J(J+1) = \sum_{k=1}^{3}\langle\Phi_{crHF}|\hat{J}_k|\Phi_{crHF}\rangle^2 = \sum_{k=1}^{3}(\hbar\Omega_k\mathcal{I}_o)^2 = (\hbar\mathcal{I}_o)^2 \cdot \sum_{k=1}^{3}\Omega_k^2 = (\hbar\mathcal{I}_o)^2 \cdot \Omega^2$$

---

[8] If we include the potential-energy cross terms, then from Eqs. (24) (or (31)), (26), and (34) we obtain (and similar expressions for $\hat{J}_2$ and $\hat{J}_3$):

$$0 = \langle\Phi_{crHF}|[\hat{H}_{crHF},\hat{J}_1]|\Phi_{crHF}\rangle = \langle\Phi_{crHF}|[\hat{H}_{oHF},\hat{J}_1]|\Phi_{crHF}\rangle + i\hbar\left(\Omega_2 \cdot \langle\Phi_{crHF}|\hat{J}_3|\Phi_{crHF}\rangle - \Omega_3 \cdot \langle\Phi_{crHF}|\hat{J}_2|\Phi_{crHF}\rangle\right)$$
$$= \langle\Phi_{crHF}|[\hat{H}_{oHF},\hat{J}_1]|\Phi_{crHF}\rangle + i\hbar\Omega_2 \cdot \Omega_3 \cdot (\bar{\mathcal{I}}_3 - \bar{\mathcal{I}}_2)$$

where the left-hand-side commutator vanishes because $|\Phi_{crHF}\rangle$ is an eigenstate of $\hat{H}_{crHF}$. Since now $x$-$y$-$z$ degrees of freedom are all coupled, $\langle\Phi_{crHF}|[\hat{H}_{oHF},\hat{J}_1]|\Phi_{crHF}\rangle$ does not vanish and hence $\bar{\mathcal{I}}_3 \neq \bar{\mathcal{I}}_2$ and from Eq. (26) we conclude that $\vec{J}$ is not parallel to $\vec{\Omega}$ and hence in general uniform rotation is not possible.

which is simplified to:
$$\sqrt{J(J+1)} = \Omega \mathscr{S}_o \qquad (59)$$
where $\mathscr{S}_o$ is defined in (55). Eq. (59) determines $\Omega$.

The HF self-consistency condition on the oscillator frequencies $\omega_k$ in Eq. (34) is expressed by the following minimization of the intrinsic energy $E_{JgscrHF}$ in Eq. (47) subject to the constant volume condition in Eq. (35) (using Feynman's theorem [71]):

$$0 = \frac{\partial}{\partial \omega_2} E_{JgscrHF} = \langle \Phi_{crHF} | \frac{\partial \hat{H}}{\partial \omega_2} | \Phi_{crHF} \rangle = \langle \Phi_{crHF} | \frac{\partial \hat{H}_o}{\partial \omega_2} | \Phi_{crHF} \rangle = -M\omega_1 \frac{\omega_o^3}{\omega_2^2 \omega_3} \langle x^2 \rangle + M\omega_2 \langle y^2 \rangle$$

and similarly for derivative with respect to $\omega_3$ yielding the conditions:

$$\omega_1^2 \langle x^2 \rangle = \omega_2^2 \langle y^2 \rangle = \omega_3^2 \langle z^2 \rangle \qquad (60)$$

where $\langle x_k^2 \rangle \equiv \langle \Phi_{crHF} | \sum_{n=1}^{A} x_{nk}^2 | \Phi_{crHF} \rangle$, which are evaluated using Eqs. (51), (52), and (53) and the definitions in Eq. (15). Combining Eqs. (35) and (60), we obtain:

$$\omega_1^2 = \omega_o^2 \cdot \left( \frac{\langle y^2 \rangle \cdot \langle z^2 \rangle}{\langle x^2 \rangle^2} \right)^{1/3}, \quad \omega_2^2 = \omega_o^2 \cdot \left( \frac{\langle x^2 \rangle \cdot \langle z^2 \rangle}{\langle y^2 \rangle^2} \right)^{1/3}, \quad \omega_3^2 = \omega_o^2 \cdot \left( \frac{\langle x^2 \rangle \cdot \langle y^2 \rangle}{\langle z^2 \rangle^2} \right)^{1/3} \qquad (61)$$

The coupled Eqs. (59), (60), (61) determine $\tilde{\Omega}_k$, $\Omega$, and $\omega_k$ as functions of $J$. In these equations $\omega_1'$, $\alpha_2'$, $\alpha_3'$, $\omega_+'$, $\omega_-'$, $\omega_2'$, $\omega_3'$, and $\gamma$ are given in terms of $\omega_k$ and $\tilde{\Omega}_k$ by Eqs. (41), (42), (43), (48), and (49). In general, these equations are solved iteratively. However, because we have ignored the potential-energy cross terms $x'y'$ and $x'z'$ in Eq. (39) to reduce it to that of a uniaxial cranking model, only one of the three equations in Eq. (58) can have a solution, and this solution describes a small-amplitude or slow angular-momentum precession. For this solution we arbitrarily choose Eq. (58) with $k = 1$, where $\tilde{\Omega}_1$ is large and $\tilde{\Omega}_2$ and $\tilde{\Omega}_3$ are very small. The other two equations in Eq. (58) are approximately well satisfied for sufficiently small $\tilde{\Omega}_2$ and $\tilde{\Omega}_3$. Eq. (58) with $k = 1$ and Eq. (53) for $\mathscr{S}_1^o$ yield:

$$\mathscr{S}_{rig} - 4 \frac{\alpha_2' \Sigma_2 - \alpha_3' \Sigma_3}{\alpha_2'^2 - \alpha_3'^2} = \left(1 - \tilde{\Omega}_1^2\right) \frac{\Sigma_1}{\omega_1'} + \frac{1+\tilde{\Omega}_1^2}{2} \mathscr{S}_{rig}$$
$$+ \frac{\omega_-'^2}{\alpha_2'^2 - \alpha_3'^2} \left( \frac{\Sigma_2}{\alpha_2'} - \frac{\Sigma_3}{\alpha_3'} \right) \cdot \left( \frac{1+\tilde{\Omega}_3^2}{1-\tilde{\Omega}_3^2} \cdot \tilde{\Omega}_1^2 - 1 \right) \cdot \frac{1}{\sqrt{1+\tan^2 2\gamma}} \qquad (62)$$
$$- \frac{\omega_-'^2}{\alpha_2'^2 - \alpha_3'^2} \left( \frac{\Sigma_2}{\alpha_2'} - \frac{\Sigma_3}{\alpha_3'} \right) \cdot \frac{2\Omega_1 \Omega_2 \Omega_2}{1-\tilde{\Omega}_3^2} \cdot \frac{\tan 2\gamma}{\sqrt{1+\tan^2 2\gamma}}$$

In absence of precession, $\tilde{\Omega}_2$ and $\tilde{\Omega}_3$ vanish, $\tilde{\Omega}_1 = 1$ (from Eq. (38)), and consequently Eq. (62) requires that $\alpha_2' \Sigma_2 - \alpha_3' \Sigma_3 = 0$. This is the isotropic-velocity-distribution condition of Bohr-Mottelson [49] and Ripka-Blaizot-Kassis [50] used in their analysis with the conventional cranking model for a rotation about a single principal axis of the potential energy. This condition together with the constant potential volume condition in Eq. (35)[9] predicted the rigid-flow kinematic moment of inertia $\mathcal{I}_{rig}$ in Eq. (50) and constant frequencies $\omega'_1$, $\alpha'_2$, and $\alpha'_3$, and hence constant intrinsic energy $E_{JgscrHF}$ in Eq. (47). For $\tilde{\Omega}_1 \neq 1$ in Eq. (62), $\alpha'_2 \Sigma_2 - \alpha'_3 \Sigma_3 \neq 0$, and we generalize the Bohr-Mottelson [49] and Ripka-Blaizot-Kassis [50] condition to the following two conditions (which satisfy Eq. (62)):

$$\mathcal{I}_{rig} - 4 \frac{\alpha'_2 \Sigma_2 - \alpha'_3 \Sigma_3}{\alpha'^2_2 - \alpha'^2_3} = \frac{1+\tilde{\Omega}_1^2}{2} \mathcal{I}_{rig} \tag{63}$$

and

$$\left(1-\tilde{\Omega}_1^2\right)\frac{\Sigma_1}{\omega'_1} + \frac{\omega'^2_-}{\alpha'^2_2 - \alpha'^2_3}\left(\frac{\Sigma_2}{\alpha'_2} - \frac{\Sigma_3}{\alpha'_3}\right)\cdot\left(\frac{1+\tilde{\Omega}_3^2}{1-\tilde{\Omega}_3^2}\cdot\tilde{\Omega}_1^2 - 1\right) = 0 \tag{64}$$

where we have ignored the last term on the right-hand-side of Eq. (62) because $\tan 2\gamma$ is very small since $\tilde{\Omega}_2$ and $\tilde{\Omega}_3$ are very small (at least for the $^{20}_{10}Ne$ case), but they can be taken into account iteratively.

Solving Eqs. (63) and (64), we obtain:

$$\tilde{\Omega}_1^2 = 1 - \frac{8}{\mathcal{I}_{rig}} \cdot \frac{\alpha'_2 \Sigma_2 - \alpha'_3 \Sigma_3}{\alpha'^2_2 - \alpha'^2_3} \tag{65}$$

$$\tilde{\Omega}_3^2 = \left(1-\tilde{\Omega}_1^2\right)\cdot\left(\frac{\Sigma_1}{\omega'_1} - \Delta\right) \Big/ \left[\frac{\Sigma_1}{\omega'_1} - \Delta - \left(\frac{\Sigma_1}{\omega'_1} + \Delta\right)\cdot\tilde{\Omega}_1^2\right] \tag{66}$$

And from Eq. (38) we obtain:

$$\tilde{\Omega}_2^2 = 1 - \tilde{\Omega}_1^2 - \tilde{\Omega}_3^2 \tag{67}$$

where:

$$\Delta \equiv \frac{\omega'^2_-}{\alpha'^2_2 - \alpha'^2_3}\left(\frac{\Sigma_2}{\alpha'_2} - \frac{\Sigma_3}{\alpha'_3}\right) \tag{68}$$

As $\Omega$ increases with $J$, we can show from Eqs. (48) and (49) that $\omega'^2_-$ decreases. At some critical values $\Omega_c$ and $J_c$ of $\Omega$ and $J$, $\omega'^2_-$ vanishes. Eq. (64) then shows that, at $\Omega_c$,

$$\tilde{\Omega}_1 = 1, \text{ i.e., } \Omega_1 = \Omega_c, \qquad \tilde{\Omega}_2 = \tilde{\Omega}_3 = 0 \tag{69}$$

and hence from Eq. (50) we obtain:

$$\langle \Phi_{crHF} | \hat{J}_1 | \Phi_{crHF} \rangle = \hbar \Omega_1 \mathcal{I}_o = \hbar \Omega_c \mathcal{I}_o = \hbar J_c, \qquad \langle \Phi_{crHF} | \hat{J}_k | \Phi_{crHF} \rangle = 0 \text{ for } k=2,3 \tag{70}$$

Eqs. (69) and (70) show that, at $J_c$, the angular velocity and the expectation of the angular momentum vectors become completely aligned with the x or 1 axis of the space-fixed frame[10] since the angles $\phi$ and $\theta$ in Eq. (36) vanish. In this aligned state, we must use the angular momentum constraint in the first of Eq. (70) instead of Eq. (59). Solving the first of Eq. (70), we obtain:

$$J_c = \Sigma_3 - \Sigma_2 \tag{71}$$

which is the familiar result from the conventional uniaxial cranking model [49,50,64,65].

---

[9] They actually used constant nuclear volume condition but this difference does not have significant consequences, refer to the footnote 6.

[10] Incidentally in this aligned state, Eq. (65) shows that $\alpha_2 \Sigma_2 - \alpha_3 \Sigma_3 = 0$ as expected for a uniaxial rotation, refer to the related statement in the paragraph immediately preceding Eq. (63).

In Eqs. (65) and (66), we determine $\omega_1'$, $\alpha_2'^2$, $\alpha_3'^2$ and $\Omega$ from solving iteratively Eqs. (42), (48), (49), (59), (60), and (61).

The excited-state ($E_{JcrHF}$) and excitation ($\Delta E_{JcrHF}$) energies of the ground-state rotational band are given by:

$$E_{JcrHF} = \langle \Phi_{crHF} | \hat{H}_{oHF} | \Phi_{crHF} \rangle = \langle \Phi_{crHF} | \hat{H}_{crHF} | \Phi_{crHF} \rangle + \vec{\Omega} \cdot \langle \Phi_{crHF} | \vec{\hat{J}} | \Phi_{crHF} \rangle$$
$$= E_{JgscrHF} + \vec{\Omega} \cdot \langle \Phi_{crHF} | \vec{\hat{J}} | \Phi_{crHF} \rangle \tag{72}$$

$$\Delta E_{JcrHF} \equiv E_{JgscrHF} - E_{J=0\,gscrHF} + \vec{\Omega} \cdot \langle \Phi_{crHF} | \vec{\hat{J}} | \Phi_{crHF} \rangle \tag{73}$$

In Eqs. (72) and (73), we use the angular momentum constraint $\sqrt{J(J+1)} = \Omega \mathcal{S}_o$ in Eq. (59) below the rotational-band termination angular momentum $J_c$ in Eq. (71), and use the constraint $J_c = \Omega_c \mathcal{S}_o$ at the band termination point at $J_c$.

For a rotating nucleus with triaxial deformation, the electric quadrupole moment is given by [49]:

$$Q = \sqrt{Q_o^2 + Q_2^2}, \quad Q_o \equiv \frac{Z}{A}e\sum_{n=1}^{A}(2z_n^2 - x_n^2 - y_n^2), \quad Q_2 \equiv \frac{Z}{A}e \cdot \sqrt{\frac{3}{2}} \cdot \sum_{n=1}^{A}(x_n^2 - y_n^2) \tag{74}$$

where $Q_o$ is the quadrupole moment for an EM transition associated with the rotation about the x-axis, and $Q_2$ is the quadrupole moment for an EM transition associated with the rotation about the y-axis, Z and A are respectively the total proton and nucleon numbers, and e the electric charge. Eqs. (74) for $Q$ is used even at $J_c$ where $\langle \Phi_{crHF} | \vec{\hat{J}} | \Phi_{crHF} \rangle$ is aligned with $\vec{\Omega}$ because there is still quantum fluctuation in the operator $\vec{\hat{J}}$ and the residual of the square of the angular momentum contribution has, so far, not been accounted for (refer to Eq. (30)).

5. **Prediction of Eqs. (59) and (65), (66), (67), (69), (70), and (71) for slow-wobbling rotation in $^{20}_{10}Ne$**

We have iteratively solved the microscopic self-consistent cranking-model (MSCRM-3W) Eqs. (59) and (65), (66), (67), (69), (70), and (71) for slow-wobbly rotation in the $^{20}_{10}Ne$ nucleus using $P = 35.43$ in $\hbar\omega_o = P.A^{-1/3}$ and using the cranked HF state $|\Phi'_{crHF}\rangle \equiv \hat{P}\prod_{n=1}^{A}|\varphi'_n\rangle$ (in Eqs. (44), (45), (46), (47), and (54)) obtained by filling each of the cranked harmonic-oscillator orbitals $|\varphi_n\rangle = U|\varphi'_{n,os}\rangle \equiv U|n_1 n_2 n_3\rangle$ with two protons and two neutrons yielding the ground-state

prolate configuration $(000)^4 (100)^4 (010)^4 (001)^4 (002)^4$ (and the orbitals $(200), (020)$, $(110), (101), (011)$ and higher-lying orbitals are unoccupied).

Fig 1 shows the MSCRM-3W predicted ground-state rotational-band excitation energy $\Delta E_{JcrHF}$ in Eq. (73) versus $J$ in $^{20}_{10}Ne$. Also shown in Fig 1 for comparison are the excitation energy from the microscopic uniaxial self-consistent cranking model (MSCRM-1) in [1], and that observed in experiments. Fig 1 shows that MSCRM-3W excitation energy is reasonably well predicted although it is consistently lower than the measured value. However, the excitation energy can be increased to match the measured energy by using $P = 38$ instead of 35.43. In contrast, MSCRM-1 highly overpredicts the excitation energy.

Of particular interest in Fig 1 is the progressive reduction in the measured rotational-band energy-level spacing with increasing $J$, originally observed by Bohr-Mottelson who attributed it to some unknown collective phenomenon [49]. This reduction is well predicted by MSCRM-3W but not predicted at all by MSCRM-1[11]. In MSCRM-3W (where $E_{JgscrHF}$ is minimized), the level spacing decreases with $J$ because the intrinsic energy $E_{JgscrHF}$ decreases with $J$ (since the frequencies $\alpha'_2$, and $\alpha'_3$ decrease) more than the rotational energy $\vec{\Omega} \cdot \langle \Phi_{crHF} | \vec{J} | \Phi_{crHF} \rangle$ increases with $J$ (refer to Eq. (72)). In MSCRM-1 [1], however, $E_{JgscrHF}$ remains constant (to satisfy the rigid-flow angular momentum constraint) and therefore the level spacing increases with $J$. The level spacing also increases with $J$ in the Ripka-Blaizot-Kassis conventional cranking model [50]. On the other hand, in other conventional uniaxial cranking models [49, 65,66,67] (where the Ripka-Blaizot-Kassis condition is not enforced and instead $E_{JgscrHF}$ is minimized) $E_{JgscrHF}$ decreases with $J$ but not sufficiently to overcome the increasing rotational energy, and therefore the level spacing increases with $J$. MSCRM-3W predicts the reduction in the level spacing with $J$ because it admits a small collective rotation transverse to that along a principal axis of the deformed nucleus, in other words it admits a small-amplitude precession or wobbling motion of the angular momentum.

This precession reduces the kinematic moment of inertia below the rigid-flow value $\mathcal{I}_{rig}$ as discussed in the statement immediately following Eq. (55). This can be seen in Eqs. (50), (58), and (63): in Eq. (63) we observe that if $\tilde{\Omega}_1 = 1$ (i.e., if $\Omega_1 = \Omega$), then we have the isotropic-velocity-distribution condition $\alpha_2 \Sigma_2 - \alpha_3 \Sigma_3 = 0$, and hence the rigid-flow constraint in Eq. (58)

---

[11] In fact, this reduction is also not predicted by other models such as HF, SU(3), Sp(3,R), and phenomenological approaches [3,59,63,64,65,67,72-81]. In [65,81], which uses self-consistent deformed oscillator with $\vec{l} \cdot \vec{s}$ coupling and without two-body interaction, the predicted excitation energy in $^{20}_{10}Ne$ follows a straight line up to $J = 6$ and is lower than the measured excitation energy by as much as 2.4 MeV at $J = 6$ and $J = 8$. The smaller predicted energy spacing between $J = 6$ and $J = 8$ relative to that between $J = 4$ and $J = 6$ is achieved by assuming that the oblate aligned state at $J = 8$ is rotating about the rotation axis, which at $J = 8$ is also the symmetry axis.

(also refer to Eq. (26)) along the principal axis. For wobbling motion, $\tilde{\Omega}_1 < 1$, $\tilde{\Omega}_2$ and $\tilde{\Omega}_3$ are non-zero but small, and $\alpha_2 \Sigma_2 - \alpha_3 \Sigma_3 \neq 0$ (except at $J = 8$), and hence the kinematic moment of inertia on the left-hand-side of Eq. (58) or (63) about the principal axis is smaller than $\mathcal{S}_{rig}$. At the cut-off angular momentum $J_c$, $\tilde{\Omega}_1 = 1$, and the kinematic moment of inertia increases to the rigid-flow moment.

In the $^{20}_{10}Ne$ ground-state rotational band predicted by MSCRM-3W, the nucleus is triaxial oblate. Therefore, the predicted wavefunction does not have good signature and angular momentum quantum numbers, and hence the wavefunction should be a superposition of odd and even angular momentum eigenstates (as discussed in [6,10]). It is surprising then that no EM (magnetic M1 and electric moment) transitions with $\Delta J = 1$ have been reported for $^{20}_{10}Ne$ (to the author's knowledge).

Fig 2 shows that the predicted electric quadrupole moment defined in Eq. (74) agrees closely with the measured quadrupole moment except at $J = 4$ where it is overpredicted by about 25%. The cause of this overprediction is not yet understood but the measured data at $J = 4$ seems suspiciously low.

We have analyzed the impact of the residuals of the square of the angular momentum and of two-body interaction in Eq. (30) on the results of MSCRM-3W in Figs 1 and 2 by solving Eq. (30) using Eq. (29) for $\left(\hat{J}_k^2\right)_{res}$, Eq. (33) for $\hat{V}_{res}$, and Tamm-Dancoff approximation similarly to that in [1]. The result of this analysis is not presented in this article because it does not significantly alter the results in Figs 1 and 2.

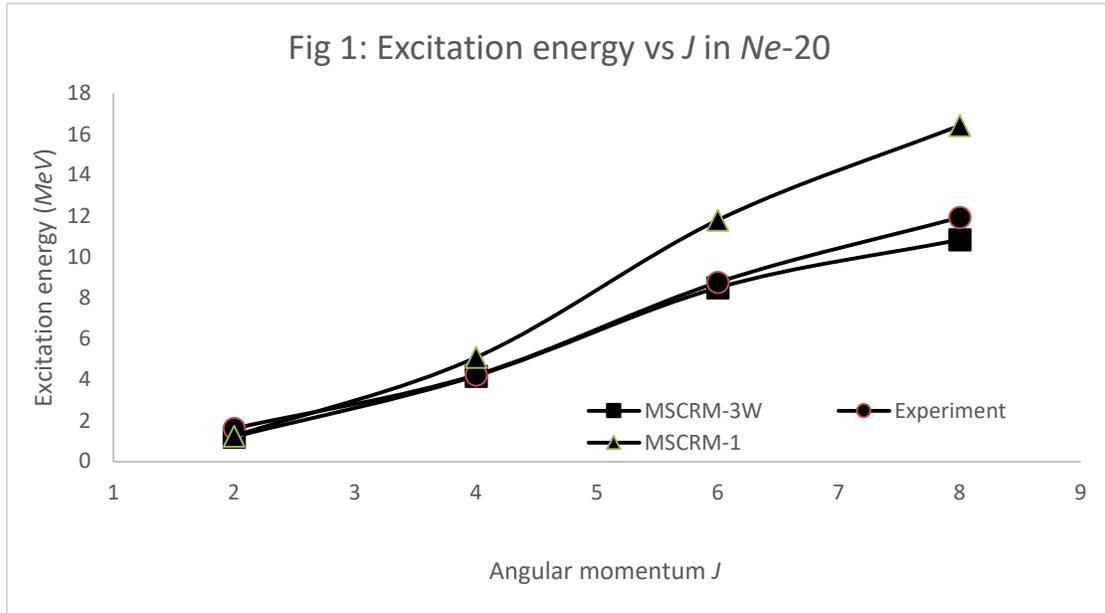

Fig 1: Excitation energy vs *J* in *Ne*-20

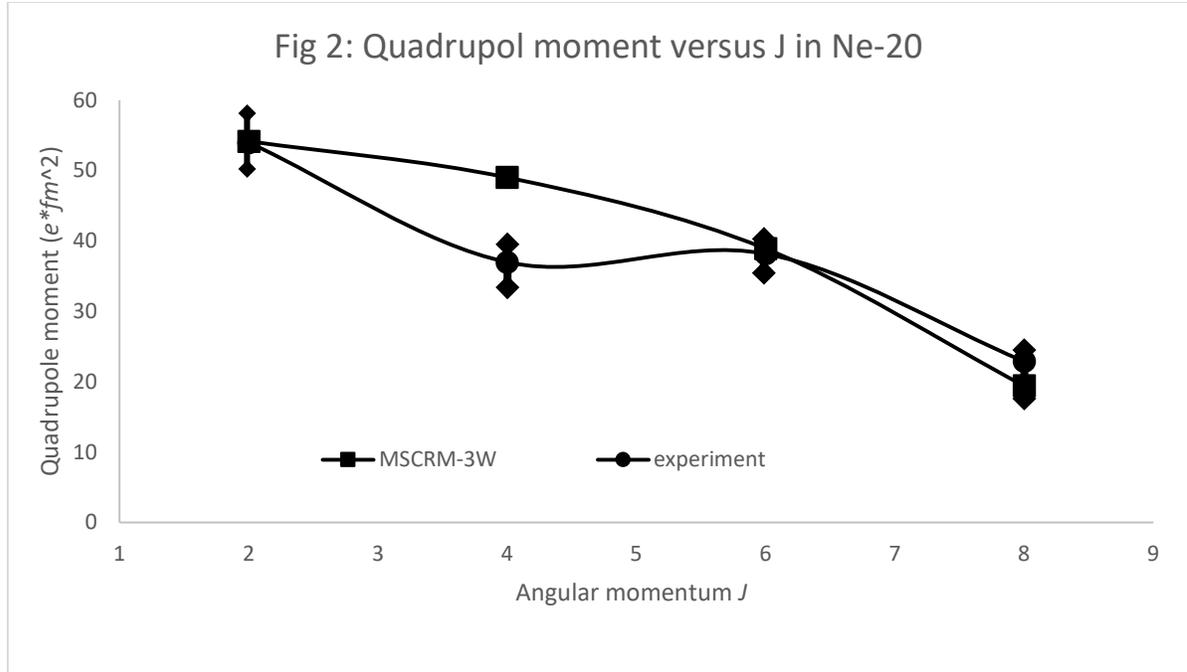

Fig 2: Quadrupol moment versus J in Ne-20

## 6. Concluding remarks

In this article, a microscopic quantum ideal rotor model intrinsic Hamiltonian for triaxial rotation is derived from the nuclear Schrodinger equation by applying a rotationally invariant rotation operator to a nuclear ground-state (or excited-band band-head) wavefunction describing a deformed nucleon distribution. The three rotation angles in the rotation operator are chosen to specify the orientation of the quadrupole component of the deformed nucleon distribution, and hence they are considered to be functions of the nucleon co-ordinates. The rotor-model Hamiltonian becomes purely (ideally) intrinsic, i.e., it becomes effectively independent of angular momentum operator, when each of the three rotation angles is chosen to describe a rigid-flow component of nucleon velocity field, and canonically conjugate to the angular momentum operator. The intrinsic Hamiltonian resembles the intrinsic Hamiltonian in the Bohr's rotational model but with the rigid-flow kinematic moment of inertia instead of an arbitrary dynamic moment of inertia. The Hamiltonian is, however, distinct from that in the Bohr's rotational model. It is argued that the ideal rotor model is more appropriate for the description of the rotational motion in at least light nuclei than the Bohr's rotational model. (For the case of the center-of-mass motion, the rotor model predicts the correct mass.)

Using Hartree-Fock and second-quantization methods, the ideal triaxial-rotor-model Hamiltonian is reduced to a self-consistent cranking-model (MSCRM-3) Hamiltonian for a rotation about a self-consistently-defined axis plus correction terms related to the residuals of the square of the angular momentum and a two-body interaction. It is shown that the conventional cranking-model Hamiltonian is not purely intrinsic because it does not use a rigid-flow kinematic moment of inertia. Unlike that in the conventional cranking model, the rotation-angle vector in MSCRM-3 is not arbitrary but given self-consistently by the ratio of the expectation of the angular momentum and rigid-flow kinetic moment of inertia. Therefore, it cannot be varied arbitrarily as in the conventional cranking model. In particular, one cannot a priori assume a uniform rotation (i.e., angular momentum vector being parallel to rotation-angle vector) except under some stringent conditions and example of which is presented belwo.

For a deformed harmonic-oscillator self-consistent potential and using a rotation of the co-ordinate system, the MSCRM-3 Schrodinger equation is transformed into that of a uniaxial cranking model MSCRM-3W (where a component of the angular momentum is parallel to the rotation-angle vector) plus two local potential-energy cross terms. For a slow-wobbling rotation about the rotation-angle vector, these potential-energy cross terms are shown to be negligibly small. The MSCRM-3W Schrodinger equation is then solved analytically using a generalization of the isotropic-velocity-distribution condition of Bohr-Mottelson and Ripka-Blaizot-Kassis.

For $^{20}_{10}Ne$, MSCRM-3W predicts a ground-state rotational-band excitation energy that agrees well with the measured excitation energy. In particular, the measured decrease in energy-level spacing with increasing angular momentum, which was originally noted by Bohr-Mottelson [49] and has so far eluded prediction, is well predicted and occurs for the following reasons. The generalized Bohr-Mottelson and Ripka-Blaizot-Kassis condition mentioned above allows the oscillator frequencies to decrease with increasing angular momentum much more than that in the conventional cranking model (including the more restrictive Ripka-Blaizot-Kassis where the frequencies remain constant). The faster decrease in the frequencies causes the intrinsic energy to decrease much more than that in the other models resulting in a decreasing rotational-band energy-level spacing with increasing angular momentum.

In the excited slow-wobbly-rotational states, the nuclear shape is predicted to be approximately axially-symmetric oblate and to become exactly axially symmetric at the band termination. Therefore, the Hamiltonian possesses $D_2$ symmetry and hence the wavefunction signature is approximately conserved. Therefore, the rotational states have approximately even angular momenta, in agreement with the measurement. MSCRM-3W predicts well the quadrupole moment including the sharp drop near the band termination in $^{20}_{10}Ne$. The residuals of the square of the angular momentum and two-body interaction are found to have insignificant impact on the MSCRM-3W results.

In a future article, we will study the solution of the full MSCRM-3 Schrodinger equation to examine whether uniform rotation can occur under other conditions, and to study the conditions and regions for tilted rotations.

**References**

[1]  P. Gulshani, arXiv:1810.11836 [nucl-th], October 28, 2018, revised January 16, 2019.
[2]  E.R. Marshalek, Nucl. Phys. A311 (1979) 429.
[3]  M.G. Vassanji and M. Harvey, Nucl. Phys. A344 (1980) 61.
[4]  F. Cuypers, Nucl. Phys. A468 (1987) 237.
[5]  H. Frisk and R. Bengtsson, Phys. Letts. B 196 (1987) 14.
[6]  S. Frauendorf, Nucl. Phys. A557 (1993) 259c.
[7]  A.F. Dodaro and A.L. Goodman, Nucl. Phys. A 596 (1995) 91.
[8]  T. Horibata and N. Onishi, Nucl. Phys. A 596 (1996) 251.

[9]   W.D. Heiss and R.G. Nazmitdinov, Phys. Letts. B 397 (1997) 1.
[10]  S. Frauendorf, Nucl. Phys. A 677 (2000) 115.
[11]  V. I. Dimitrov, S. Frauendorf, and F. Dönau, Phys. Rev. Lett. 84 (2000) 5732.
[12]  S. Frauendorf, Rev. Mod. Phys. 73 (2001) 463.
[13]  W.D. Heiss and R.G. Nazmitdinov, Phys. Rev. C 65 (2002) 054304.
[14]  I. Hamamoto, Phys. Rev. C 65 (2002) 044305-1.
[15]  M. Matsuzaki, Y.R. Shimizu, and K. Matsuyanagi, Phys. Rev. C 65 (2002) 041303(R)-1.
[16]  M. Oi and P.M. Walker, Phys. Letts. B 576 (2003) 75.
[17]  M. Matsuzaki and S-I. Ohtsubo, Phys. Rev. C 69 (2004) 064317-1.
[18]  M. Matsuzaki, Y.R. Shimizu, and K. Matsuyanagi, Phys. Rev. C 69 (2004) 034325-1.
[19]  D.J. Thouless and J.G. Valatin, Nucl. Phys. 31 (1962) 211.
[20]  A. Faessler, W. Greiner, and R.K. Sheline, Nucl. Phys. 70 (1965) 33.
[21]  A. Faessler, W. Greiner, and R.K. Sheline, Nucl. Phys. 80 (1965) 417.
[22]  J.M. Eisenberg and W. Greiner, Nuclear Theory, (North Holland, Amsterdam, 1970).
[23]  H.J. Mang, Physics Report 18, no. 6 (1975) 325.
[24]  P. Ring and P. Schuck, The Nuclear Many-Body Problem (Springer-Verlag, N.Y., 1980).
[25]  A.K. Klein, Phys. Rev. C 63 (2000) 014316.
[26]  F.M.H. Villars, Nucl. Phys. 74 (1965) 353.
[27]  R.E. Peierls and J. Yaccoz, Proc. Phys. Soc. 70 (1957) 381.
[28]  F. Villars and N. Schmeing-Rogerson, Ann. Phys. 63 (1971) 443.
[29]  F. Villars and G. Cooper, Ann. Phys. (N.Y.) 56 (1970) 224.
[30]  P. Gulshani, Nucl. Phys. A 832 (2010) 18.
[31]  P. Gulshani, Nucl. Phys. A 852 (2011) 109.
[32]  J. Sau, J. Phys. A: Math. Gen. 12 (1979) 1971.
[33]  P. Gulshani, arXiv:1609.04267 [nucl-th], September 19, 2016.
[34]  P. Gulshani, arXiv:1610.08337 [nucl-th], September 10, 2017.
[35]  P. Gulshani, arXiv:1708.03326 [nucl-th], August 10, 2017.
[36]  Chi-Yu Hu, Nucl. Phys. 66 (1965) 449.
[37]  E.B. Bal'butsev and I.N. Mikhailov, Akad. Nauk. SSSR, Izv. Ser. Fiz. 30 (1966) 1118.
[38]  B.L. Birbrair, Nucl. Phys. A257 (1976) 445.
[39]  I.N. Mikhailov and D. Jansen, Phys. Letts. 72B (1978) 303.
[40]  D. Janssen, I.N. Mikhailov, R.G. Nazmitdinov, B. Nerlo-Pomorska, K. Pomorska, and R.Kh. Safarov, Phys. Letts. 79B (1978) 347.
[41]  E.R. Marshalek and J. Weneser, Ann. Phys. 53 (1969) 564.
[42]  B.L. Birbrair, Sov. J. Nucl. Phys. 28 (1979) 631.
[43]  A.V. Ignatyuk and I.N. Mikhailov, Sov. J. Nucl. Phys. 30 (1979) 343.
[44]  D. Janssen and I.N. Mikhailov, Nucl. Phys. A318 (1979) 390.
[45]  I.N. Mikhailov and D. Janssen, Izvestiya Adademii Nauk SSSR, Seriya Fizicheskaya, 41 (1977) 35.
[46]  K. Goeke and P.-G. Reinhard, Ann. Phys. 124 (1980) 249.
[47]  A.K. Kerman and N. Onishi, Nucl. Phys. A361 (1981) 179.

[48]  N. Onishi, Nucl. Phys. A456 (1986) 279.
[49]  A. Bohr and B.R. Mottelson, Nuclear Structure, Vol. II (Benjamin, N.Y., 1975).
[50]  G. Ripka, J.P. Blaizot, and N. Kassis, in Heavy-Ion, High-Spin States and Nuclear Structure, Vol. 1, Trieste Int. Seminar on Nuclear Physics, September 17-December 21, 1973 (IAEA, Vienna, 1975).
[51]  M.E. Rose, Elementary Theory of Angular Momentum, (John Wiley & Sons, Inc., N.Y., 1974).
[52]  A. deShalit and H. Feshbach, Theoretical Nuclear Physics, Vol. 1 (John Wiley & Sons, Inc., N.Y., 1957).
[53]  F. Villars and N. Schmeing-Rogerson, Ann. Phys. 63 (1971) 443.
[54]  A. Bohr and B.R. Mottelson, Nuclear Structure, Vol. I (Benjamin, N.Y., 1975).
[55]  B.L. Birbrair, Phys. Letts. 72B (1978) 425.
[56]  H. Goldstein, Classical Mechanics (Addison-Wesley Publishing Co. Inc., Reading Massachusetts, 1965).
[57]  T.H.R. Skyrme, Proc. Roy. Soc. (London) A70 (1957) 433.
[58]  C.A. Levinson, Phys. Rev. 132 (1963) 2184.
[59]  I. Kelson and C.A. Levinson, Phys. Rev. 134 (1964) B269.
[60]  M.K. Banerjee, D. D'Oliveira, and G.J. Stephenson, Jr., Phys. Rev. 181 (1969) 1404.
[61]  J.P. Elliott and J.A. Evans, Nucl. Phys. A324 (1979) 12.
[62]  A. Bohr, Rev. Mod. Phys. 48 (1976) 365.
[63]  M. Bouten, M.C. Bouten, and E. Caurier, Nucl. Phys. A193 (1972) 49.
[64]  T. Troudet and R. Arvieu, Ann. Phys. 134 (1981) 1.
[65]  S.G. Nilsson and I. Ragnarsson, Shapes and shells in nuclear structure (Cambridge University Press, Cambridge, UK. 1995).
[66]  J.G. Valatin, Proc. Roy. Soc. (London) 238 (1956) 132.
[67]  A.P. Stamp, Z. Physik A 284 (1978) 312.
[68]  P. Gulshani and A. B. Volkov, J. Phys. G: Nucl. Phys. 6 (1980) 1335.
[69]  G. Rosensteel, Phys. Rev. C 46 (1992) 1818.
[70]  P. Gulshani, arXiv [nucl-th] 1602.08337, November 03, 2016.
[71]  R.P. Feyman, Phys. Rev. 56 (1939) 340.
[72]  W.H. Bassichis, B. Giraud, and G. Ripka, Phys. Rev. Letts. 15 (1965) 980.
[73]  I. Kelson, Phys. Rev. 160 (1967) 775.
[74]  Y. Abgrall, G. Baron, E. Caurier, and G. Monsonego, Nucl. Phys. A131 (1969) 609.
[75]  M. Harvey, in Adv. in Nucl. Phys., editors M. Baranger and E. Vogt, Vol 1 (Plenum Press, N.Y. 1968).
[76]  A.P. Stamp, Nucl. Phys. A161 (1971) 81.
[77]  Y. Abgrall, B. Morand, and E. Caurier, Nucl. Phys. A192 (1972) 372.
[78]  E. Caurier and B. Grammaticos, Nucl. Phys. A279 (1977) 333.
[79]  B. Grammaticos and K.F. Liu, Il Nuovo Cimento 50 A (1979) 349.
[80]  J.P. Draayer, K.J. Weeks, and G. Rosensteel, Nucl. Phys. A413 (1984) 215.
[81]  I. Ragnarsson, S. Aberg, and R.K. Sheline, Physica Scripta 24 (1981) 215.